\documentclass[superscriptaddress,twocolumn,prb,showpacs,preprintnumbers,amsmath,amssymb]{revtex4}
\usepackage{graphicx,afterpage}
\begin{document}

\title{$^{63}$Cu-NMR study of oxygen disorder in ortho-II YBa$_2$Cu$_3$O$_y$}

\author{T. Wu}
\altaffiliation[Present address: ]{Hefei National Laboratory for Physical Sciences at the Microscale, University of Science and Technology of China (USTC), Anhui, Hefei 230026, P. R. China.}
\affiliation{Laboratoire National des Champs Magnétiques Intenses, CNRS - Universit\'e Grenoble Alpes - EMFL, 38042 Grenoble, France}
\author{R. Zhou}
\affiliation{Laboratoire National des Champs Magnétiques Intenses, CNRS - Universit\'e Grenoble Alpes - EMFL, 38042 Grenoble, France}
\author{M. Hirata}
\affiliation{Laboratoire National des Champs Magnétiques Intenses, CNRS - Universit\'e Grenoble Alpes - EMFL, 38042 Grenoble, France}
\author{I. Vinograd}
\affiliation{Laboratoire National des Champs Magnétiques Intenses, CNRS - Universit\'e Grenoble Alpes - EMFL, 38042 Grenoble, France}
\author{H. Mayaffre}
\affiliation{Laboratoire National des Champs Magnétiques Intenses, CNRS - Universit\'e Grenoble Alpes - EMFL, 38042 Grenoble, France}
\author{R. Liang}
\affiliation{Department of Physics and Astronomy, University of British Columbia, Vancouver, BC, Canada, V6T~1Z1}
\affiliation{Canadian Institute for Advanced Research, Toronto, Canada}
\author{W.~N.~Hardy}
\affiliation{Department of Physics and Astronomy, University of British Columbia, Vancouver, BC, Canada, V6T~1Z1}
\affiliation{Canadian Institute for Advanced Research, Toronto, Canada}
\author{D.~A.~Bonn}
\affiliation{Department of Physics and Astronomy, University of British Columbia, Vancouver, BC, Canada, V6T~1Z1}
\affiliation{Canadian Institute for Advanced Research, Toronto, Canada}
\author{T. Loew}
\affiliation{Max-Planck-Institut for Solid State Research, Heisenbergstra{\ss}e 1, D-70569 Stuttgart, Germany} 
\author{J. Porras}
\affiliation{Max-Planck-Institut for Solid State Research, Heisenbergstra{\ss}e 1, D-70569 Stuttgart, Germany} 
\author{D. Haug}
\affiliation{Max-Planck-Institut for Solid State Research, Heisenbergstra{\ss}e 1, D-70569 Stuttgart, Germany} 
\author{C.T.~Lin}
\affiliation{Max-Planck-Institut for Solid State Research, Heisenbergstra{\ss}e 1, D-70569 Stuttgart, Germany} 
\author{V. Hinkov}
\altaffiliation[Present address: ]{Physikalisches Institut und R\"ontgen Center for Complex Materials Systems, Universit\"at W\"urzburg, 97074 W\"urzburg, Germany.}
\affiliation{Max-Planck-Institut for Solid State Research, Heisenbergstra{\ss}e 1, D-70569 Stuttgart, Germany} 
\author{B. Keimer}
\affiliation{Max-Planck-Institut for Solid State Research, Heisenbergstra{\ss}e 1, D-70569 Stuttgart, Germany} 
\author{M.-H. Julien}
\email{marc-henri.julien@lncmi.cnrs.fr}
\affiliation{Laboratoire National des Champs Magnétiques Intenses, CNRS - Universit\'e Grenoble Alpes - EMFL, 38042 Grenoble, France}

\date{\today}


\begin{abstract}

We show that $^{63}$Cu NMR spectra place strong constraints on both the nature and the concentration of oxygen defects in ortho-II YBa$_2$Cu$_3$O$_y$. Systematic deviation from ideal ortho-II order is revealed by the presence of inequivalent Cu sites in either full or empty chains. The results can be explained by two kinds of defects: oxygen clustering into additional chains, or fragments thereof, most likely present at all concentrations ($6.4 < y < 6.6$) and oxygen vacancies randomly distributed in the full chains for $y<6.50$ only. Furthermore, the remarkable reproducibility of the spectra in different samples with optimal ortho-II order ($y \simeq 6.55$) shows that chain-oxygen disorder, known to limit electronic coherence, is ineluctable because it is inherent to these compounds. 

\end{abstract}
\maketitle

\section{Introduction}

Dopants play a pivotal role in many correlated electron systems. Their concentration determines the electronic density, the small changes of which can dramatically modify the ground state and the physical properties.~\cite{Dagotto05,Morosan12} At the same time, dopants may impact on electronic properties via lattice strain or electrostatic effects. For instance, the random distribution of dopants produces nanoscale electronic inhomogeneity in some of the high $T_c$ cuprates~\cite{Cren01,McElroy05,Singer02,Hofman13}, chemical ordering of Pr suppresses large-scale electronic phase separation in maganites~\cite{Zhu16} or charge ordering in CoO$_2$ layers intertwines with ordering of the mobile Na$^+$ ions in Na$_x$CoO$_2$.~\cite{Julien08,Alloul09a,Young13} 

\afterpage{
\begin{table*}[]
\caption{Main characteristics of the ortho-II single crystals studied in this work. The hole doping level $p$ was determined using the calibration in Ref.~\cite{Liang06} with the values of the superconducting transition temperature $T_c$ observed in the bulk magnetization. For the $T_c=35$~K crystal, $p$ is determined from its $c$-axis lattice parameter~\cite{Liang06,Haug10} since a small concentration of impurities is suspected to reduce $T_c$. Indeed, recent crystals of improved purity have a higher $T_c$ with the same oxygen content or the same $c$-axis parameter~\cite{Blanco14}. For the crystals substituted with $^{17}$O, $p$ was determined after correcting the measured $T_c$ value by +1 K in order to account for the isotope effect on $T_c$~\cite{Wu13a}. }
\begin{ruledtabular}
\begin{tabular}{cccccl}
Oxygen		&	Oxygen		&	Hole			&	$T_c$	&		Enriched	&	Reference \\
content $y$	&	content $y$	&	doping $p$	&	 (K)		&		with		&	\\
(nominal) 		&	from NMR 	&	(hole/Cu)		&			&		$^{17}$O	&	\\
\hline
$y=6.56$ 		&	6.55			&	0.109		&	59.8		&		Yes		&	Refs.~\cite{Wu13a,Wu14}	 \\
$y=6.55$ 		&	6.55			&	0.109		&	61.3		&		No		&	Refs.~\cite{Blanco14}	 \\
$y=6.47$ 		&	6.49			&	0.088 		& 	52.3		&		Yes		&	Refs.~\cite{Hirata14}	\\
$y=6.45$ 		&	6.44			&	0.079		& 	35.0		&		No		&	Refs.~\cite{Baek12,Wu13b,Hinkov08,Haug09,Haug10}	 \\

\end{tabular}
\end{ruledtabular}
\end{table*}
\begin{figure*}[]
\centerline{\includegraphics[width=7.5in]{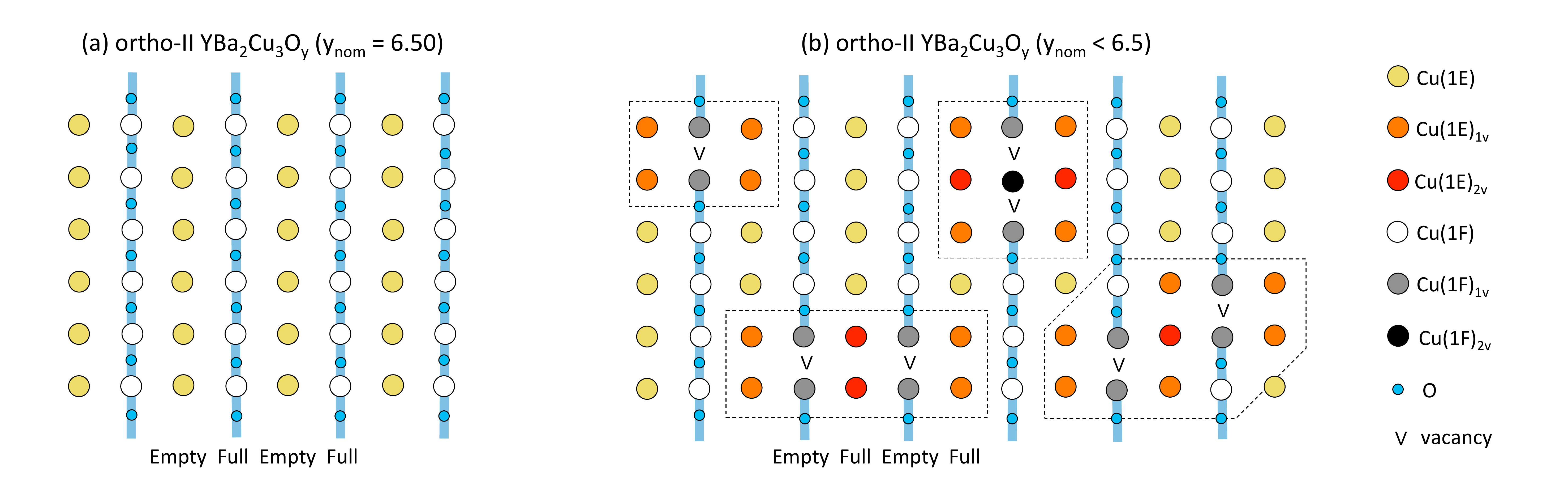}} 
\vspace{-0.1cm}
\caption{(Color online). (a) O, Cu(1E) and Cu(1F) chain sites with ideal ortho-II order. (b) Differentiation of oxygen-empty Cu(1E) and oxygen-filled Cu(1F) chain sites according to the number of nearest-neighbor oxygen vacancies in oxygen-deficient ortho-II. Low-probability configurations with more than 2 nearest-neighbor vacancies are not shown. Sites labelled Cu(1E) and Cu(1F) have no vacancy as nearest-neighbor and are often quoted in the literature as Cu(1)$_2$ (twofold coordinated) and Cu(1)$_4$ (fourfold coordinated), respectively, because of the presence of apical oxygen sites just above and below each of them. Cu(1F)$_{\rm 1V}$ sites are often quoted as Cu(1)$_3$ (threefold coordinated).}
\end{figure*}
}

In the high temperature superconductor YBa$_2$Cu$_3$O$_y$ with $6.3 \leq y \leq 6.6$, O dopants can be made to order into different sequences of oxygen-full and oxygen-empty chains. This makes YBa$_2$Cu$_3$O$_y$ by far the cleanest family of high-$T_c$ cuprates and thus a model system. The so-called ortho-II order in which each full chain is adjacent to two empty chains (Fig.~1a) is of considerable interest because it is the simplest and the best-ordered structure.~\cite{Andersen99,Zimmermann03,Casalta96,Liang00,Liang06,Liang12} The particularly low level of disorder that can be achieved in the ortho-II phase has played a pivotal role in the discoveries of Fermi-surface reconstruction~\cite{Doiron07} and charge order~\cite{Wu11}. 

This, however, does not render the question of the concentration and the nature of the oxygen defects illegitimate: first, since variations of $\Delta y \sim0.1$ in the oxygen concentration may produce substantial changes in the electronic properties, $y$ needs to be accurately measured and controlled, especially when comparing measurements performed on crystals from different groups or from different batches. Second, reducing $y$ below the theoretically-ideal ortho-II composition $y=6.50$ not only changes the hole doping but also inevitably introduces oxygen vacancies, thus disorder. Disentangling these two effects is likely to be important. This is particularly true near $y\simeq 6.4$ where sudden changes in the electronic properties~\cite{LeBoeuf11,Sun04,Vignolle12,Sebastian10,Rullier08,Li08,Grissonanche14} give clues on the interplay between different electronic orders~\cite{Hinkov08,Haug09,Haug10,muSR,Baek12,Blanco14,Hucker14,Wu13a,Wu13b,Hirata14,Baledent11}. Finally, there is also evidence that oxygen disorder has an impact on the phenomenology of charge ordering~\cite{Wu14,LeTacon14} and more broadly on quasiparticle scattering~\cite{Bobowski10}, even in the best ortho-II samples. Therefore, clarifying the concentration and type of oxygen defects in the chain layer remains an important task, even (and perhaps especially) in clean YBa$_2$Cu$_3$O$_y$ samples. 

In this article, we use NMR spectroscopy to study the different Cu sites in the ortho-II structure of YBa$_2$Cu$_3$O$_y$: Cu(1F) sites in the oxygen-full chains, Cu(1E) sites in the empty chains (see Fig~1a) as well as planar Cu(2F) and Cu(2E) sites, which are connected to Cu(1F) and Cu(1E), respectively, via an apical oxygen. The observation that the NMR signal associated with each of these crystallographic sites actually consists of several inequivalent sites provides us with microscopic information on the concentration and the spatial distribution of oxygen defects in the chain layer.

\section{NMR methods}

\subsection{Experimental details}

The main characteristics of the high quality, untwinned, single crystals studied here are summarized in table I. 

$^{63}$Cu NMR spectra were obtained by sweeping the frequency at a fixed magnetic field value ($H\simeq15$~T, aligned along the crystalline $c$-axis) and by adding the Fourier transforms of the spin-echo signal recorded for regularly-spaced frequency values. Measurement of the spin-echo decay as a function of the time $\tau$ between $\pi/2$ and $\pi$ pulses showed a Gaussian-type decay for all Cu(1E) peaks, with $T_2$ values ranging from 900~$\mu$s (left peak in Fig.~\ref{comparison}b) to 650~$\mu$s (right peak). Therefore, for Cu(1E) spectra measured with $\tau\simeq 20$~$\mu$s, the $T_2$ correction to the intensity is entirely negligible. For Cu(1F), on the other hand, $T_2$ corrections are necessary (see later).

\subsection{Background}

Any local change in the spatial configuration of oxygen ions in the chain layer is susceptible to affect the NMR resonance frequency of the nearby sites. This site differentiation may arise from two channels: the magnetic hyperfine interaction and the electric quadrupole interaction. 

The magnetic hyperfine interaction (between nuclear and electronic spins) is responsible for a shift $K$ of the three lines for each $^{63}$Cu site. Recall that because $^{63}$Cu has a nuclear spin $I=3/2$, the NMR spectra consist of a central line ($m_I$=-1/2 to 1/2 transition) and two "quadrupole" satellites ($m_I$=-3/2 to -1/2 and $m_I$=+1/2 to +3/2 transitions). See Refs.~\cite{Wu11,Wu13b} for examples of full $^{63}$Cu NMR spectra. In the cuprates, $^{63}K$ is determined by the spin susceptibility on a given nuclear site as well as on its nearest neighbors (dipolar coupling to more distant sites is negligible).

The electric quadrupole interaction is the interaction between the (quadrupole moment of the) nucleus and the electric-field-gradient (EFG). When it is a perturbation with respect to the Zeeman interaction (which is the case here at 15 T), any EFG change at a given site modifies the position of the central line of this site (only to second order) as well as the separation ($\nu_{\rm sat-sat}$) between its quadrupole satellites (first-order correction): 
\begin{equation}
\nu_{\rm sat-sat}=\frac{eQV_{zz}}{2h}(3\cos^2 \theta - 1 - \eta\sin^2 \theta \cos2\phi),
\end{equation}
where $Q$ is the electric quadrupole moment of the nucleus, $V_{zz}\equiv\partial^{2}V/\partial z^{2}$ is the largest component of the electric-field-gradient (EFG) tensor ($V$ is the electrostatic potential), $\theta$ and $\phi$ are the polar and azimuthal angles of the applied magnetic field with respect to the principal axes of the EFG tensor and $\eta$ is the asymmetry parameter of the EFG tensor~\cite{EFG}. 

The difference in EFG created by the presence or the absence of oxygen is mainly what makes Cu(1E) to be distinguishable from Cu(1F) and Cu(2E) to be distinguishable from Cu(2F) in the NMR spectra of ortho-II samples.~\cite{Wu11,Yamani04}

\subsection{Expected oxygen defects}

We expect that the main source of defects in our samples lies in the deviation from the theoretically-ideal ortho-II concentration $y=6.50$. On general grounds, the same interactions that favor oxygen order should work against having lone, random, defects. Some indication of oxygen clustering has indeed been obtained in ortho-I samples.~\cite{cluster1} For $y > 6.50$, clustering of the extra $\delta y =y-6.50$ oxygens into full chains (or fragments thereof) is to be expected from the manifest propensity to systematically order at higher oxygen concentrations. Indeed, new ordered structures (of period-8, 5 or 3) are stabilized for $\geq 6.6$~\cite{Liang12}. 

For $y < 6.50$, on the other hand, clustering of the $\delta y =6.50-y$ vacancies into empty chains may not be significant since ortho-II order never transforms into another ordered phase on decreasing $y$. Lone, random, vacancies may thus be dominant for $y < 6.50$. 

Note that the interstitial position in-between chains is unoccupied and that phase separation, while possible in principle, will be shown to be inconsistent with our results. In all cases, O vacancies should affect the EFG at both Cu(1E) and Cu(1F) sites nearby and thus have signatures in the NMR spectra.

\section{Results}

\subsection{Overview of the results}

The main results of this study are shown in Figs. 2-4. Basically, we find that the ideal ortho-II structure (namely, single Cu(1E), Cu(1F), Cu(2E) and Cu(2F) sites) is observed in {\it none} of our samples. Specifically:

$\bullet$ All samples, both $y<6.50$ and $y>6.50$, show several Cu(1F) sites (Fig.~\ref{Cu1F}) that might be related to the existence of (fragments of) extra full chains. 

$\bullet$ For $y\simeq6.55$, the deviations from ideal ortho-II order turn out to be extremely reproducible from one sample to another (Fig.~\ref{comparison}).

$\bullet$ Crystals with $y<6.50$ show several inequivalent Cu(1E) sites (Fig.~\ref{Cu1E}), which can be explained by the presence of lone oxygen vacancies in filled chains.

In the following, we discuss these results in more detail.

\subsection{Extra full-chain fragments for all concentrations}

\begin{figure}[t!]
\centerline{\includegraphics[width=2.9in]{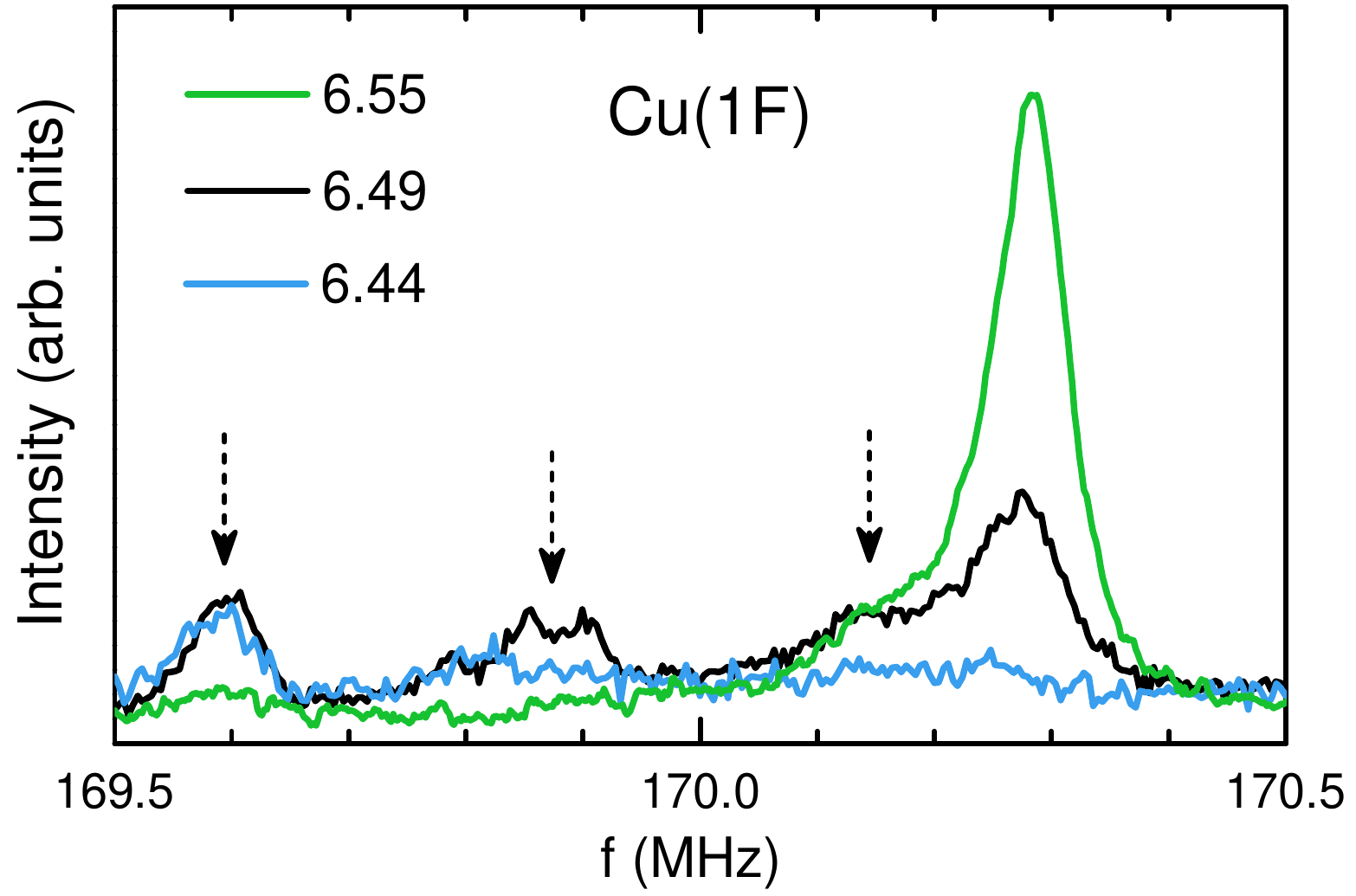}} 
 \caption{(Color online). $^{63}$Cu(1F) central line for three different values of the oxygen concentration $y$ at $T=220$~K. Besides the main site near 170.3~MHz, at least three other sites are observed on the low frequency side (marked by arrows). Intensities have been normalized by the sample weight but not by $T_2$ (which, for any given peak, hardly varies between the three samples). The main peak has shorter $T_2$ than the side peaks but if intensities are renormalized by $T_2$ values.}
\label{Cu1F}
\end{figure}

\begin{figure*}[t!]
\centerline{\includegraphics[width=5.8in]{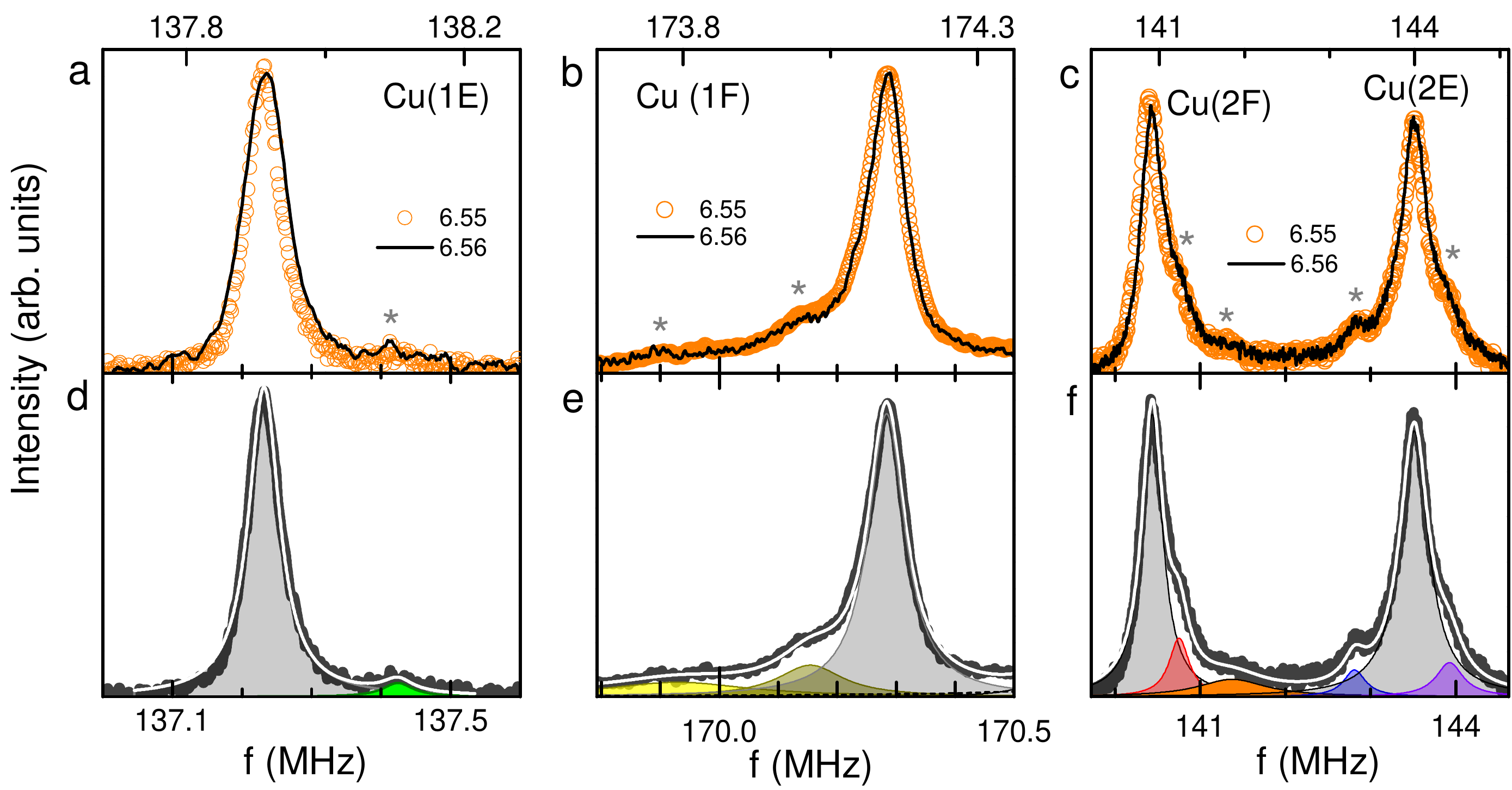}} 
\caption{(Color online). Comparison of NMR spectra of two different single crystals grown in Vancouver and Stuttgart: low-frequency satellite of Cu(1E) (a), central line of Cu(1F) (b), low-frequency satellite of Cu(2F) and Cu(2E) (c). All the secondary peaks or shoulders (marked by asterisks) are due to deviations from the ideal ortho-II structure and they prove perfectly reproducible. d,e,f: decomposition of the spectra into main sites (grey) and secondary sites (colored). The white trace represents the sum of the main and secondary sites.}
\label{comparison}
\end{figure*}

Let us first consider the concentration $y=6.55$ for which a non-zero NMR intensity is visible on the low-frequency side of the regular Cu(1F) central line (Fig.~\ref{Cu1F} and \ref{comparison}b,e). The main difference between these additional sites and the main line resides in their Knight shift value~\cite{CommentCu(1F)}. 

In a previous study, similar Cu(1F) NMR signal was initially ascribed to sites near the chain ends~\cite{Yamani06}. Indeed, because oxygen vacancies break up full chains into fragments, chain ends are expected to experience charge (Friedel) and spin-density oscillations that decay away from a vacancy, along the chain~\cite{Alloul09}. However, a subsequent theoretical work~\cite{Chen09} ruled out this interpretation (such oscillations do exist but they only contribute to the symmetric broadening of the main line) and instead proposed that the additional signal actually corresponds to a different valence state, namely Cu$^{+}$ ions. However, our spectra in Fig.~\ref{Cu1F} show that this anomalous intensity consists of not one but at least three different peaks in addition to the main line. Furthermore, all these peaks broaden and shift on cooling (not shown), indicating that none of them can be ascribed to a non-magnetic valence configuration. 

In principle, insight into the origin of the different peaks can be gained from their relative areas. Here, however, we cannot reliably estimate the relative abundance of the different sites as a fraction of the Cu(1F) sites appears not to be observed, even in the $y=6.55$ sample. This can be seen from Fig.~\ref{Cu1F} that shows small Cu(1F) peaks at the very same positions in the $y_{\rm nominal}=6.47$ and 6.45 samples as in the 6.55 sample. It is clear that the main peak almost entirely looses its intensity upon decreasing $y$. This is because some of the sites have $T_2$ values shorter than the spectrometer dead time. The loss of intensity is clear for the 6.45 and 6.47 samples by comparison with the 6.55 sample, but it is also present, to a lesser extent, in the 6.55 sample itself. Therefore, the relative abundance of the different Cu(1F) sites cannot be reliably extracted, even in the $y=6.55$ sample.

Nevertheless, it should be clear that the additional Cu(1F) lines cannot correspond to sites near lone vacancies in full-chains or near lone extra oxygen in empty chains. First, the relative intensities of the side peaks (which all have similar $T_2$ values longer than $T_2$ of the main peak) do not match the calculated number of Cu(1F) sites having $N=0, 1, 2$ random vacancies as nearest neighbors (Fig.~\ref{cu2}). Second, such lone defects should be visible in Cu(1E) NMR, which is not the case in the 6.55 sample to within our detection limit of $\sim$1\% (see Fig.~\ref{Cu1E}). The only anomalous signal found in the Cu(1E) spectra of the 6.55 sample is a small peak (asterisk in Fig.~\ref{comparison}a) that amounts to $\sim$4\% of the main peak intensity. Its origin is undetermined yet; it could be related to the anomalous Cu(1F) peaks.

A natural candidate to explain extra Cu(1F) peaks is oxygen "over-stoichiometry". Indeed, diffraction studies have demonstrated that ortho-II order is actually strongest ({\it i.e.} has largest amplitude and longest correlation length) for $y\simeq6.55$ rather than for the theoretically optimal value $y=6.50$.~\cite{Zimmermann03,Casalta96,Blanco14,Pekker91} We know that our crystal has optimal ortho-II order, and therefore has this particular $y\simeq6.55$ concentration, for several reasons: 1) it shows the narrowest NMR line widths of all the samples we have investigated, 2) analysis of the signal intensity is consistent with $y\sim6.55$ (ref.~\cite{intensityratio}), 3) crystals from the same batch show particularly strong quantum oscillations as well as particularly sharp transitions in sound-velocity measurements~\cite{privatecomm}, and 4) these features are perfectly reproducible in several crystals with $y=6.55-6.56$ (we shall return to this point below). 

We conclude that the additional Cu(1F) sites most likely reside in extra full chains (which makes them essentially invisible from Cu(1E) NMR), even though we neither have a clear picture nor an unambiguous site assignment. We speculate that the different NMR sites could correspond to fragments of different lengths. In the $y=6.55$ sample, these extra full-chains naturally account for the extra 0.05 oxygen needed to achieve optimal ordering with respect to the ideal 6.50 concentration. Nonetheless, these extra fragments of full chains turn out to be present in samples with $y<6.50$ as well. We shall show below that Cu(1E) NMR data also (indirectly) support the hypothesis of extra full chains.

\subsection{Inherent oxygen disorder}

The $y=6.56$ crystal was grown in Vancouver while the $y=6.55$ crystal was grown in Stuttgart. Nonetheless, as Fig.~\ref{comparison} shows, all deviations of the lineshapes from a simple Gaussian or Lorentzian, even the subtlest ones, are identical in the two samples. This is equally true for Cu(1E) and Cu(1F) as well as for Cu(2E) and Cu(2F) signals. This observation is perfectly consistent with the idea of well-defined, systematic, thus "intrinsic", oxygen disorder near $y=6.55$. 

Actually, this particular oxygen disorder is very likely to be present throughout the phase diagram of underdoped YBa$_2$Cu$_3$O$_y$, whatever the ordered phase: it is known that ortho-VIII and ortho-III orders are stabilized for $y\simeq6.67$ and $y\simeq6.75$, respectively, rather than for the expected values $y\simeq6.625$ and 6.67. This again suggests the presence of "extra" 0.05-0.08 oxygen in the chains. 

This result implies that oxygen disorder will always limit the CDW correlation length or the electronic mean free path, whatever the quality of the samples.

\subsection{Lone oxygen vacancies for $y<6.50$}

\begin{figure}[t!]
\centerline{\includegraphics[width=3.1in]{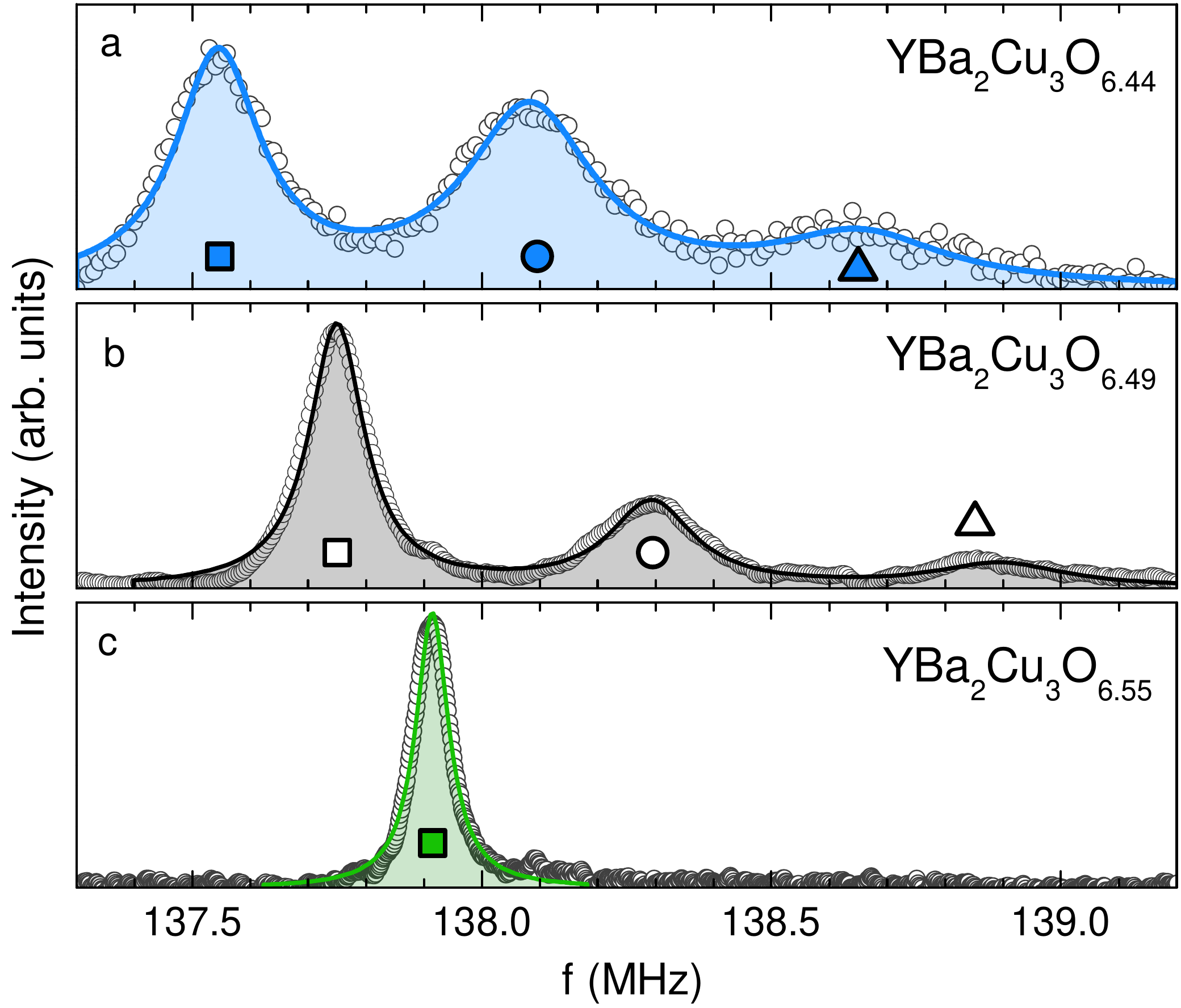}} 
\caption{(Color online). $^{63}$Cu(1E) low frequency quadrupole satellites~\cite{CommentCu(1E)} at $T=16$~K for $y=6.44$ and $T=60$~K for $y=6.49$ and 6.55 (no temperature dependence of the relative intensities was found, to within experimental uncertainty). Lines are fits with Lorentzian functions. Fits with Gaussian functions give virtually identical results.}
\label{Cu1E}
\end{figure}

For crystals with $y<6.50$, different Cu(1E) sites are observed on the quadrupole satellites (Fig.~\ref{Cu1E}), not on the central lines. This indicates that these sites mostly differ by their quadrupolar parameters (quadrupole frequency $\nu_Q$ and/or to asymmetry parameter $\eta$) rather than by their Knight shifts. Given that the positions of the additional Cu(1E) satellites correspond to relatively minor changes of $\nu_{\rm sat-sat}$ with respect to the main line (representing Cu(1E) far from any defect), we assume that the principal axis of their EFG tensor lies along the $c$-axis, as for the main site, namely $\theta=0$ in Eqn.~1. Therefore, the differentiation seen here arises from distinct values of $\nu_{\rm sat-sat}=\frac{eQV_{zz}}{h}=2\nu_Q$ and no information regarding possible differences of $\eta$ values can be gained from the spectra. 

\begin{figure}[t!]
\centerline{\includegraphics[width=2.9in]{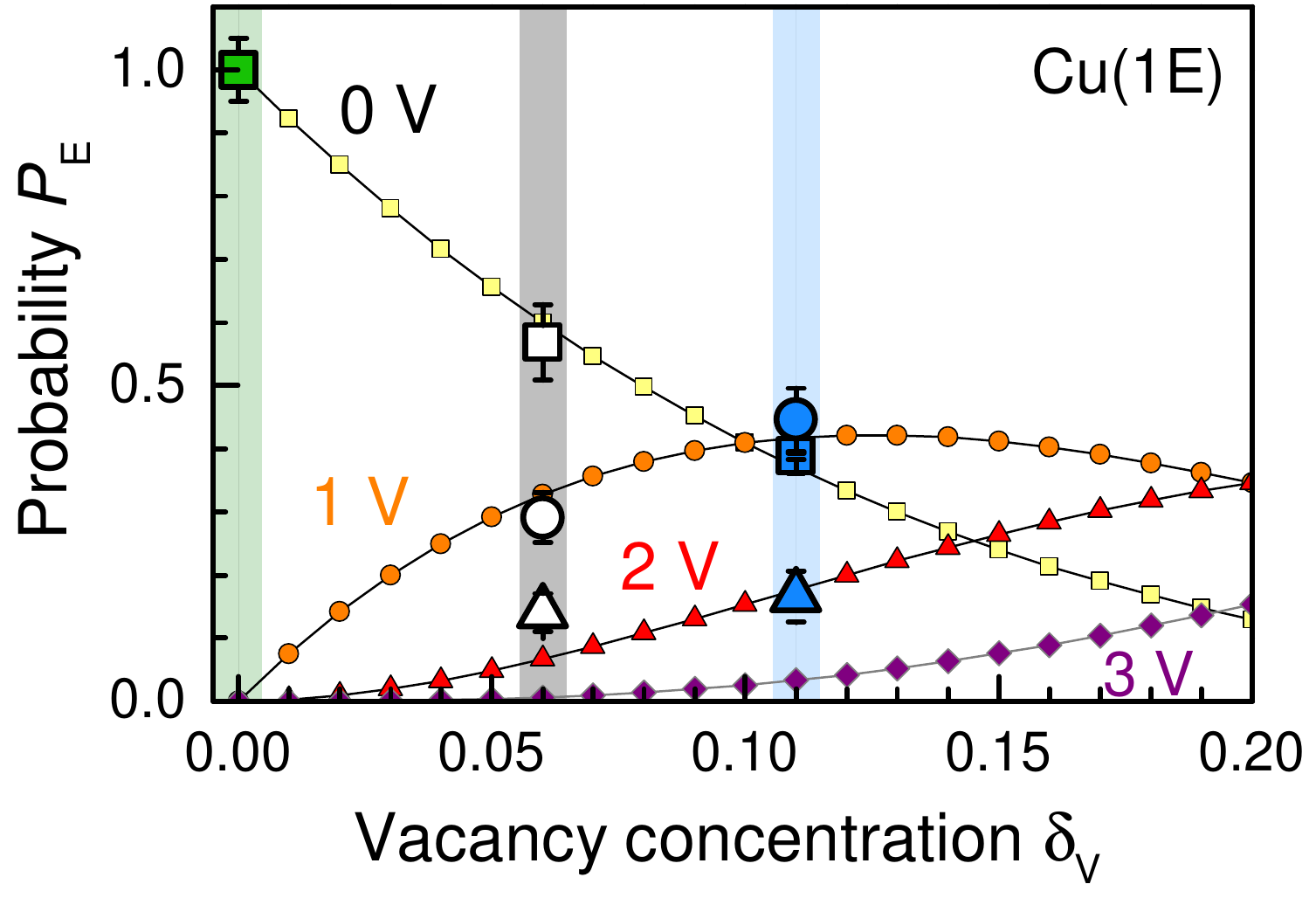}} 
 \caption{(Color online). Probability for a Cu(1E) site to have $N=0,1,2,3$ vacancies (V) among its four nearest-neighbors as a function of vacancy concentration $\delta_{\rm V}$ (per formula unit). The colors follow those of the different Cu(1E) sites in Fig.~1.  The large colored symbols represent the relative integrated intensities of the Cu(1E) NMR lines for the three samples (the symbol shapes and colors follow those of the three Cu(1E) lines in the spectra of Fig.~\ref{Cu1E}). }
 \label{probabilities}
\end{figure}

As mentioned above, vacancies in the ortho-II pattern are to be expected for $y<6.50$. Fig.~\ref{probabilities} shows the calculated probabilities
$P_E(N,\delta_{\rm V})=\left(^4_N \right)(2\delta_{\rm V})^N (1-2\delta_{\rm V})^{4-N}$ \( \)
for a Cu(1E) site to have $N=0$ to 3 randomly-distributed vacancies (V) among its four nearest neighbors as a function of vacancy concentration $\delta_{\rm V}=6.5-y$ per total formula unit (that is, per total number of chain-Cu site). A good matching between this calculation and the relative areas of the Cu(1E) peaks is unequivocally obtained if the vacancy concentration is $\delta_{\rm V}\simeq 0.06$ for the $p=0.088$ sample (nominally 6.45) and $\delta_{\rm V}\simeq 0.11$ for $p=0.079$ (nominally 6.47). 

If we assume that the same extra 0.05 oxygen present for $y=6.55$ are also present at lower doping (see above), the above-determined vacancy concentrations lead to the actual formulas ($y=6.55-\delta_{\rm V}$): YBa$_2$Cu$_3$O$_{6.49}$ for $p=0.088$ and YBa$_2$Cu$_3$O$_{6.44}$ for $p=0.079$. These oxygen contents are consistent with expectations from the $T_c$ values of the samples and from their nominal oxygen content, to within an uncertainty of about $\pm0.02$. Using $y=6.50-\delta_{\rm V}$, on the other hand, leads to less consistent numbers.

It should be noted that there exist different configurations producing Cu(1E) sites with two nearest-neighbor vacancies (see Fig.~1b) and that there is no {\it a priori} reason for the modifications of the EFG tensor to be identical for these different Cu(1E)$_{\rm 2V}$ sites (Cu(1E) sites with two nearest-neighbor vacancies). Nonetheless, we observe that the change in $\nu_{Q}$ is a relatively small effect ($\Delta\nu_{Q}/\nu_{Q}\simeq1.7$~\% between the first ($\square$) and second ($\bigcirc$) peaks), and is twice larger between the first and third ($\triangle$) peaks (see appendix). This suggests that the changes in $\nu_{Q}$ are, to first approximation, proportional to the number of nearest-neighbor vacancies, regardless of their location. 

Previous NMR studies for $y<6.5$ reported very similar Cu(1E) NMR spectra (\cite{Lutgemeier92,Poulakis00} and references therein). However, the $\delta_{\rm V}$ vacancies were assumed to order and to create additional empty chains. Accordingly, the different Cu(1E) sites were assigned to the three following configurations: empty chains surrounded by two full chains (pristine ortho-II), pairs of adjacent empty chains and sets of three adjacent empty chains. However, the Cu(1E) and Cu(2) line intensities (Fig.~\ref{probabilities}) calculated in such an hypothesis would require a significantly lower oxygen content to be consistent with the data. Therefore, generalized clustering into very long chains appears to be inconsistent with the quantitative success of the model of random vacancies. Furthermore, the values of the linewidth and of the NMR relaxation times $T_1$ and $T_2$ of Cu(1F) sites in the $y=6.55$ sample are notably different from those in the two samples with more vacancies~\cite{Wu14b}. This is consistent with random vacancies significantly reducing the average length of the full-chain fragments. Some oxygen clustering may be present but this is not what differentiates the Cu(1E) sites in NMR and it is not crucial for determining the vacancy concentration in the full chains.

Our data in the doping range $y=6.44-6.55$ also appear inconsistent with the disordered oxygen-poor domains proposed to be intertwined with ortho-II "puddles" in YBa$_2$Cu$_3$O$_{6.33}$~\cite{Campi13}: if significant parts of the sample had no oxygen order, the random-distribution model would not be quantitatively consistent with the data and the sites from disordered regions should be detected in NMR, which is not the case. 

In conclusion, the consistency between modeling and experimental results leads us to conclude that the different Cu(1E) sites observed in the NMR spectra of ortho-II YBa$_2$Cu$_3$O$_y$ with $y<6.5$ correspond to different positions with respect to randomly-distributed oxygen vacancies in the full chains. In appendix, we show that the hierarchy of $\nu_{Q}$ values for the different sites are consistent with this assignment (Fig.~\ref{nuq}) and that the random distribution of lone vacancies is also consistent with data from the planar Cu(2) sites (Fig.~\ref{cu2}).

\section{Perspectives and conclusions}

Chain-Cu NMR constitutes an interesting, non-destructive, method to probe oxygen defects in ortho-II YBa$_2$Cu$_3$O$_y$. In the future, it would be interesting to study ordering/disordering of the chains~\cite{Achkar14} by monitoring the Cu(1E/F) NMR spectrum as a function of pressure and/or annealing time. It would also be interesting to search for similar site differentiation in $^{17}$O NMR. 

Studying other ordered phases, such as ortho-VIII and ortho-III, could provide valuable information but is likely to be more challenging because their shorter-ranged oxygen order leads to broader lines and their more complex oxygen patterns generate a larger number of inequivalent NMR sites. 

To summarize, our NMR spectra provide direct evidence that chain-oxygen disorder is both ineluctable and intrinsic in single crystals of ortho-II YBa$_2$Cu$_3$O$_y$. A fraction of the oxygens likely clusters into full-chain fragments at all concentrations ($6.4 < y < 6.6$) while oxygen vacancies randomly distributed in the full chains are identified only for $y<6.50$. For optimal ortho-II order ($y \simeq 6.55$), the reproducibility of the spectra shows that chain-oxygen disorder is ineluctable because it is inherent to these compounds. 

\section*{Acknowledgments}

We are indebted to M. Horvati\'c and F. Lalibert\'e for comments on the manuscript and to J. Baglo, W. Chen, P. Hirschfeld, C. Proust, D. Le Boeuf for discussions. This work was supported by P\^ole SMIng - Universit\'e J. Fourier - Grenoble and by the French Agence Nationale de la Recherche (ANR) under reference AF-12-BS04-0012-01 (Superfield). Part of this work was performed at the LNCMI, member of the European Magnetic Field Laboratory (EMFL). Work in Vancouver was supported by the Canadian Institute for Advanced Research and the Natural Science and Engineering Research Council. 

\section*{Appendix A: $\nu_Q$ and width values for the different lines}

\begin{figure}[b!]
\centerline{\includegraphics[width=3.in]{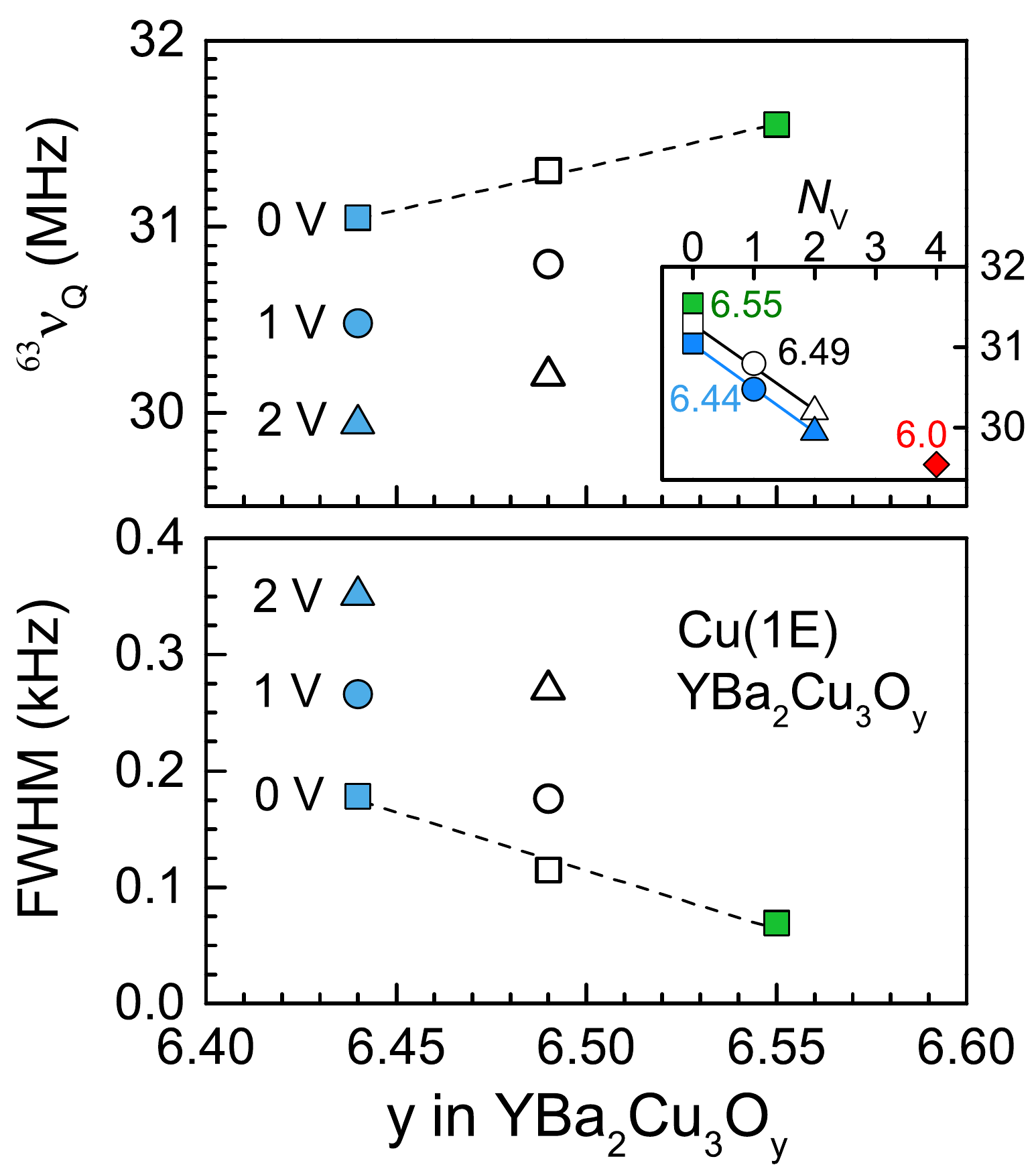}} 
 \caption{(Color online). (a) $\nu_{Q}$ as a function of the number $N_V$ of nearest-neighbor vacancies (same symbols as in Figs.~\ref{Cu1E} and \ref{probabilities}). Error bars are $\sim$0.05~MHz, less than the symbol size. Data for $y=6.0$ are from Ref.~\cite{Mali91}. (b) Full-width-at-half-maximum of the different lines.}
 \label{nuq}
\end{figure}

Experimental values of the quadrupole frequency $\nu_{Q}$ for the different lines are shown in Fig.~\ref{nuq}a at three different oxygen concentrations, as a function of the number $N$ of nearest-neighbor vacancies. Exact values of $\nu_{Q}$ in increasing order of nearest-neighbor-vacancy number are 31.05, 30.48 and 29.94~MHz for the $y=6.44$ sample, 31.34, 30.79 and 30.19~MHz for $y=6.49$ and 31.55~MHz for $y=6.55$. Note that Yamani {\it et al. } found a larger value of 32.1~MHz consistent with a higher $T_c$ and an actual oxygen content $y\simeq6.6$, in spite of the the nominal $y=6.50$.~\cite{Yamani04} The decrease of $\nu_{Q}$ with increasing vacancy number is consistent with the decrease of $\nu_{Q}$ of Cu(1E) as the oxygen content is decreased from pure ortho-II order (0 nearest-neighbor vacancies) to the undoped material ($y=6.0$, 4 nearest-neighbor vacancies). 

Width data in Fig.~\ref{nuq}b show that more underdoped samples are more inhomogeneous, presumably due to stronger chemical inhomogeneity.

\section*{Appendix B: Cu(1F), Cu(2) and Cu(2F) intensities for random vacancies}

\begin{figure}
\centerline{\includegraphics[width=3.in]{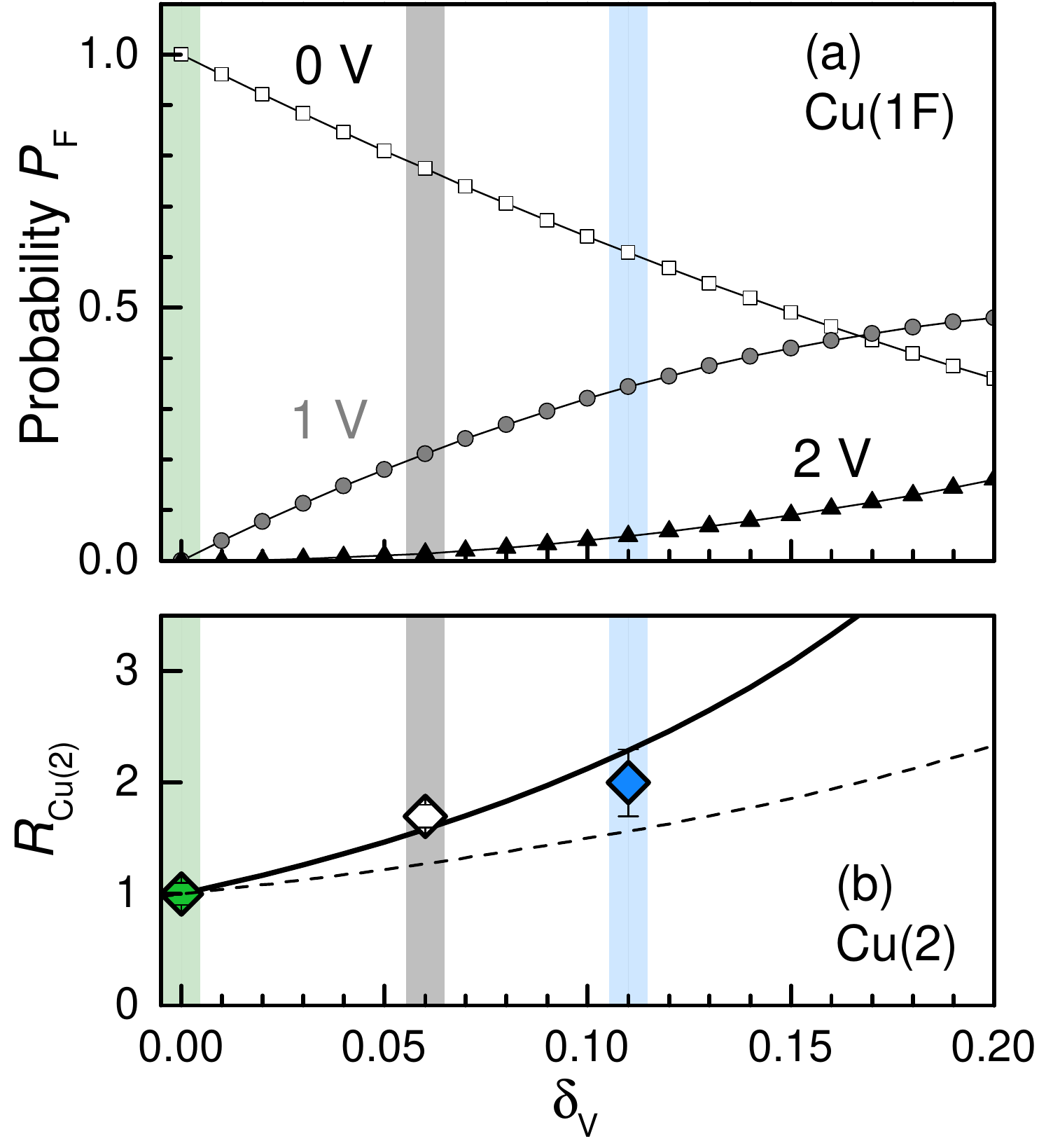}} 
 \caption{(Color online). (a) Probability for a Cu(1F) site to have $N=0,1,2$ vacancies among its two nearest-neighbors. Symbol colors follow those of Cu(1F) sites in Fig.~1. (b) Solid curve: ratio $R_{\rm Cu(2)}=I_{\rm Cu(2E)}/I_{\rm Cu(2F)}$ under the assumption that the planar Cu(2F) sites close to one, or more, vacancy resonate in the planar Cu(2E) line. Symbols: experimental values. Dashed curve: calculated ratio if all of the vacancies form empty chains.}
 \label{cu2}
\end{figure}

Fig.~\ref{cu2}a shows the probability $P_F(N,\delta_{\rm V})=\left(^2_N\right) (2\delta_{\rm V})^N (1-2\delta_{\rm V})^{2-N}$ for a Cu(1F) site to have $N=0,1,2$ vacancies among its two nearest-neighbors. The numbers are clearly inconsistent with the relative intensities of the different Cu(1F) lines in Fig.~\ref{Cu1F} (even considering uncertainties related to the partial wipeout of the main peak), showing that these lines cannot be attributed to the sites neighboring lone, random oxygen vacancies.

Fig.~\ref{cu2}b shows the predicted ratio $R_{\rm Cu(2)}=I_{\rm Cu(2E)}/I_{\rm Cu(2F)}$ of the integrated intensities of the "Cu(2E)" and "Cu(2F)" NMR lines, under the assumption that $\nu_{Q}$ of each of the two Cu(2F) which are nearest-neighbors to (at least) one vacancy (Fig.~1b) is closer to $\nu_{Q}$ of Cu(2E) sites than to $\nu_{Q}$ of Cu(2F) sites of the pristine ortho-II structure. Hence, these particular Cu(2F) nuclei actually resonate in the "Cu(2E)" NMR line (Fig.~\ref{comparison}f shows the composite nature of the Cu(2E) and Cu(2F) lines). The number of such sites is readily calculated  since it is equal to the number of Cu(1F) neighboring one, or more, vacancy (panel (a)). The random distribution model, with the above assumption, predicts a growth of the intensity ratio of the Cu(2E)-like NMR line at the expense of the Cu(2F)-like line with increasing number of vacancies which is again in quantitative agreement with the experimental values (uncorrected for possible $T_2$ differences). In contrast, the calculated ratio $R=(1+2\delta_{\rm V})/(1-2\delta_{\rm V})$ under the assumption that all of the vacancies form empty chains is twice smaller.


\begin{references}
\bibitem{CommentCu(1F)} We have checked this point by comparing central lines of $^{63}$Cu and $^{65}$Cu isotopes. Note that the quadrupole satellites are extremely broad and weak. They show some structure but no correspondance with the various central lines could be established.
\bibitem{intensityratio} After integration of the NMR signal of the central lines over frequency and correction for $T_2$ decay, we find that Cu(1F) is indeed more intense than Cu(1E). The intensity ratio translates into an actual oxygen content $y=6.53$, which must be considered as a lower bound because a small part of the Cu(1F) intensity is most likely wiped out in this sample.
\bibitem{CommentCu(1E)} Only low-frequency satellites are shown since at 15~T the high-frequency satellites overlap with the planar Cu(2) satellites. An example of a full $^{63}$Cu spectrum is shown in Fig.~1 of Ref.~\cite{Wu13b} (note that the coarse frequency resolution of the Cu(1E) satellite in Ref.~\cite{Wu13b} results in a misrepresentation of the relative intensity ratio of the different Cu(1E) lines).
\end{references}
\end{document}